\documentstyle[12pt,axodraw,epsf]{article} 
\begin{document}

\begin{flushright}
   CERN-TH/99-291\\[-0.2cm]
   THES-TP/99-12\\[-0.2cm]
   hep-ph/9909485\\[-0.2cm]
   September 1999
\end{flushright}

\begin{center}
{\LARGE {\bf Higgs-Boson Two-Loop Contributions to \\[0.3cm] 
Electric Dipole Moments in  the MSSM}}\\[1.5cm]
{\large Apostolos Pilaftsis}\\[0.4cm]
{\em Theory Division, CERN, CH-1211 Geneva 23, Switzerland\\[0.2cm]
and\\[0.2cm]
Department of Theoretical Physics, University of Thessaloniki,\\
GR 54006 Thessaloniki, Greece}
\end{center}
\bigskip\bigskip\bigskip \centerline{\bf ABSTRACT} 
The   complete set of Higgs-boson  two-loop  contributions to electric
dipole  moments of  the  electron and  neutron  is calculated  in  the
minimal supersymmetric  standard model.   The electric  dipole moments
are induced  by CP-violating trilinear  couplings  of the `CP-odd' and
charged Higgs bosons  to the scalar  top and bottom quarks.  Numerical
estimates of the individual  two-loop contributions to electric dipole
moments are given.
%PACS numbers: 11.30.Er, 14.80.Er

\newpage

Supersymmetric (SUSY)  theories, including the  minimal supersymmetric
standard model (MSSM), face the difficulty of explaining naturally the
apparent absence of electric dipole  moments (EDMs) of the neutron and
electron  \cite{EFN,DDLPD,PN,IN,CKP}.  Several  suggestions  have been
made to suppress the SUSY  contributions to electron and neutron EDMs,
at  a level  just  below their  present  experimental $2\sigma$  upper
bounds: $|d_e|  < 0.5 \times 10^{-26}  e$ cm and $|d_n|  < 1.12 \times
10^{-25}  e$  cm  \cite{PDG}.    Apart  from  the  obvious  choice  of
suppressing the new CP-violating phases of the theory to the $10^{-3}$
level   \cite{EFN,DDLPD},    a   more   phenomenologically   appealing
possibility is to make the first two generations of scalar fermions as
heavy as  few TeV, but keep  the soft-breaking mass  parameters of the
third generation  relatively small, e.g.\ 0.5--0.7  TeV \cite{PN}.  An
interesting alternative is to  arrange for partial cancellations among
the  different EDM  contributions \cite{IN},  within the  framework of
superstring-derived models \cite{BEKL}.

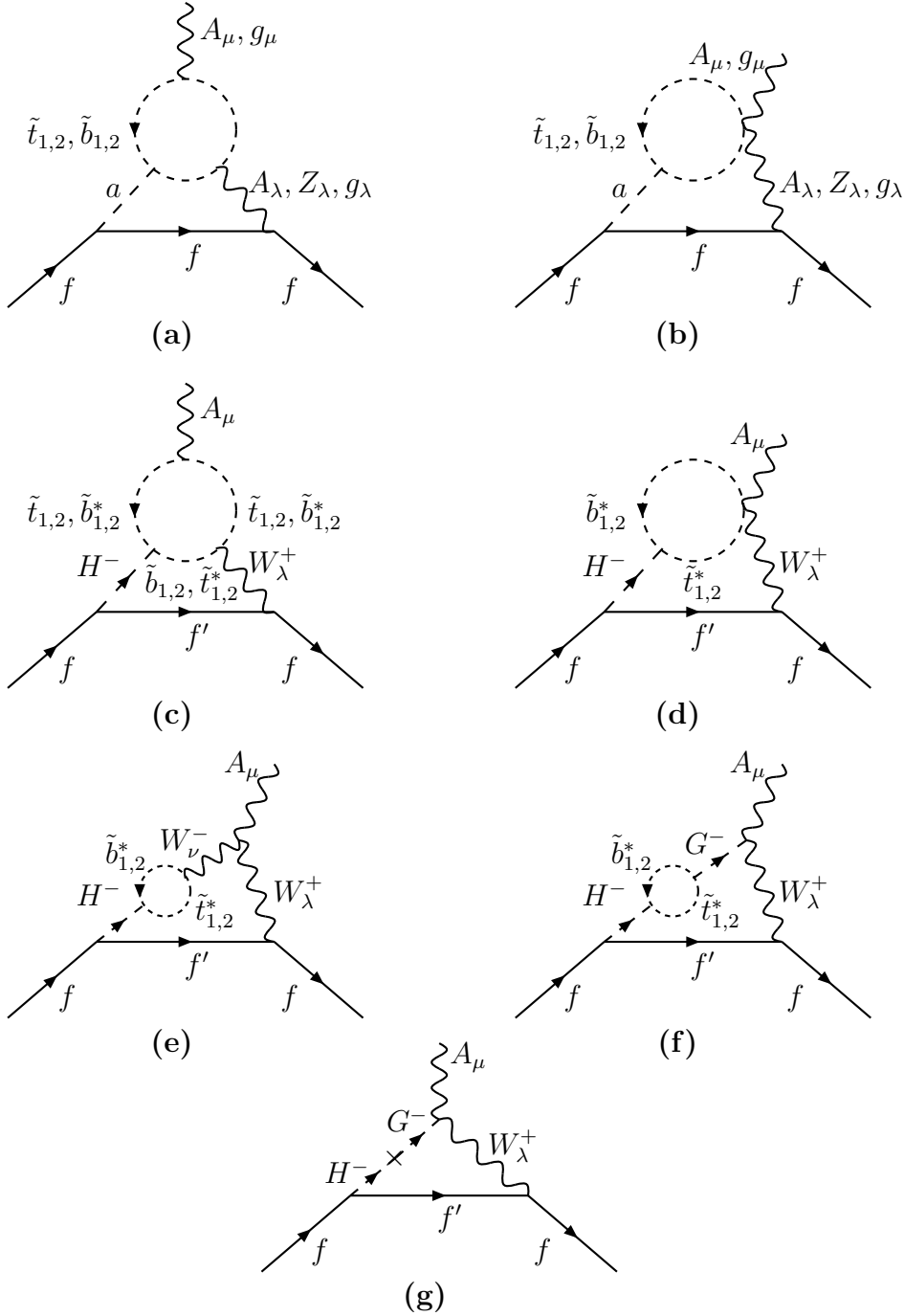
\begin{figure}[t]

\begin{center}
\begin{picture}(400,500)(0,0)
\SetWidth{0.8}
 
\ArrowLine(10,380)(45,410)\Text(30,385)[lb]{$f$}
\DashLine(45,410)(68,435){5}\Text(55,425)[rb]{$a$}
\Photon(92,435)(115,410){3}{3}
\Text(105,430)[l]{$A_\lambda,Z_\lambda,g_\lambda$}
\ArrowLine(45,410)(115,410)\Text(80,401)[l]{$f$}
\ArrowLine(115,410)(150,380)\Text(118,385)[lb]{$f$}
\Photon(80,470)(80,500){3}{3}\Text(86,490)[l]{$A_\mu,g_\mu$}
\DashArrowArc(80,450)(20,0,360){3}\Text(55,450)[r]{$\tilde{t}_{1,2},
\tilde{b}_{1,2}$} 

\Text(75,370)[]{\bf (a)}

\ArrowLine(210,380)(245,410)\Text(230,385)[lb]{$f$}
\DashLine(245,410)(268,435){5}\Text(255,425)[rb]{$a$}
\Photon(301,450)(315,410){3}{4}
\Text(315,430)[l]{$A_\lambda,Z_\lambda,g_\lambda$}
\Photon(301,450)(315,480){3}{3}\Text(310,480)[r]{$A_\mu,g_\mu$}
\ArrowLine(245,410)(315,410)\Text(280,401)[l]{$f$}
\ArrowLine(315,410)(350,380)\Text(318,385)[lb]{$f$}
\DashArrowArc(280,450)(20,0,360){3}
\Text(255,450)[r]{$\tilde{t}_{1,2}, \tilde{b}_{1,2}$}

\Text(275,370)[]{\bf (b)}

\ArrowLine(10,230)(45,260)\Text(30,235)[lb]{$f$}
\DashArrowLine(45,260)(68,285){5}\Text(55,275)[rb]{$H^-$}
\Photon(92,285)(115,260){3}{3}\Text(105,280)[l]{$W^+_\lambda$}
\ArrowLine(45,260)(115,260)\Text(80,251)[l]{$f'$}
\ArrowLine(115,260)(150,230)\Text(118,235)[lb]{$f$}
\Photon(80,320)(80,350){3}{3}\Text(86,340)[l]{$A_\mu$}
\DashArrowArc(80,300)(20,0,360){3}\Text(55,300)[r]{$\tilde{t}_{1,2},
\tilde{b}^*_{1,2}$} \Text(105,300)[l]{$\tilde{t}_{1,2},
\tilde{b}^*_{1,2}$}
\Text(83,272)[]{$\tilde{b}_{1,2},\tilde{t}^*_{1,2}$}

\Text(75,220)[]{\bf (c)}

\ArrowLine(210,230)(245,260)\Text(230,235)[lb]{$f$}
\DashArrowLine(245,260)(268,285){5}\Text(255,275)[rb]{$H^-$}
\Photon(301,300)(315,260){3}{4}\Text(315,280)[l]{$W^+_\lambda$}
\Photon(301,300)(315,330){3}{3}\Text(310,330)[r]{$A_\mu$}
\ArrowLine(245,260)(315,260)\Text(280,251)[l]{$f'$}
\ArrowLine(315,260)(350,230)\Text(318,235)[lb]{$f$}
\DashArrowArc(280,300)(20,0,360){3}
\Text(255,300)[r]{$\tilde{b}^*_{1,2}$} 
\Text(285,272)[]{$\tilde{t}^*_{1,2}$} 

\Text(275,220)[]{\bf (d)}

\ArrowLine(10,100)(45,130)\Text(30,105)[lb]{$f$}
\DashArrowLine(45,130)(63,144){4}\Text(55,145)[rb]{$H^-$}
\Photon(80,155)(101,170){3}{3}\Text(80,170)[]{$W^-_\nu$}
\Photon(101,170)(115,130){3}{4}\Text(115,150)[l]{$W^+_\lambda$}
\Photon(101,170)(115,200){3}{3}\Text(110,200)[r]{$A_\mu$}
\ArrowLine(45,130)(115,130)\Text(80,121)[l]{$f'$}
\ArrowLine(115,130)(150,100)\Text(118,105)[lb]{$f$}
\DashArrowArc(72,150)(10,0,360){2}
\Text(65,165)[r]{$\tilde{b}^*_{1,2}$} 
\Text(92,142)[]{$\tilde{t}^*_{1,2}$} 

\Text(75,90)[]{\bf (e)}

\ArrowLine(210,100)(245,130)\Text(230,105)[lb]{$f$}
\DashArrowLine(245,130)(263,144){4}\Text(255,145)[rb]{$H^-$}
\DashArrowLine(280,155)(301,170){4}\Text(286,170)[]{$G^-$}
\Photon(301,170)(315,130){3}{4}\Text(315,150)[l]{$W^+_\lambda$}
\Photon(301,170)(315,200){3}{3}\Text(310,200)[r]{$A_\mu$}
\ArrowLine(245,130)(315,130)\Text(280,121)[l]{$f'$}
\ArrowLine(315,130)(350,100)\Text(318,105)[lb]{$f$}
\DashArrowArc(272,150)(10,0,360){2}
\Text(265,165)[r]{$\tilde{b}^*_{1,2}$} 
\Text(292,142)[]{$\tilde{t}^*_{1,2}$} 

\Text(275,90)[]{\bf (f)}

\ArrowLine(110,0)(145,30)\Text(130,5)[lb]{$f$}
\DashArrowLine(145,30)(163,45){4}\Text(153,35)[rb]{$H^-$}
\DashArrowLine(163,45)(180,60){4}\Text(168,60)[]{$G^-$}
\Text(163,45)[]{\bf $\times$}
\Photon(180,60)(215,30){3}{3.5}\Text(200,50)[l]{$W^+_\lambda$}
\Photon(180,60)(180,90){3}{3}\Text(185,85)[l]{$A_\mu$}
\ArrowLine(145,30)(215,30)\Text(180,21)[l]{$f'$}
\ArrowLine(215,30)(250,0)\Text(218,5)[lb]{$f$}

\Text(175,-10)[]{\bf (g)}

\end{picture}
\end{center}
\caption{Higgs-boson two-loop contributions to EDM and CEDM of a
fermion in the MSSM in the Feynman--'t Hooft gauge (mirror graphs are
not shown); $f'$ represents the conjugate fermion of $f$ under
$T^f_z$.}\label{f1}
\end{figure}

Here,  we shall  focus our  interest on  studying  additional two-loop
contributions  to electron  and neutron  EDMs in  a SUSY  scenario, in
which the first  two generations are rather heavy,  e.g.\ of order few
TeV, whereas  the third generation  is relatively light below  the TeV
scale \cite{GD}.  In such a theoretical framework, the leading effects
on the EDMs arise  from CP-violating trilinear interactions related to
the  scalar top  and bottom  quarks through  the  three-gluon operator
\cite{DDLPD}  and through the  coupling of  the `CP-odd'  Higgs boson,
$a$,  to the  gauge bosons  \cite{CKP}, as  shown in  Fig.\  \ref{f1}. 
Despite their similarity to the  graphs due to Barr and Zee \cite{BZ},
our  reasoning  of considering  the  two-loop  contributions of  Fig.\ 
\ref{f1}   is   completely    different   from   \cite{BZ},   as   the
third-generation  scalar  quarks  can  have a  significant  impact  by
themselves  on the  electron and  neutron EDMs,  independently  of the
chirality-unsuppressed  one-loop contributions.   In our  analysis, we
shall  assume  rather   heavy  gluino  masses  $m_{\tilde{g}}$,  e.g.\ 
$m_{\tilde{g}}  >  0.5$  TeV,  such  that  their  effect  through  the
three-gluon    operator,   which    scales    as   $1/m^3_{\tilde{g}}$
\cite{DDLPD}, will be much smaller than the present experimental upper
bound \cite{IN}.

In this paper,  we shall complement the analysis  of Ref.\ \cite{CKP},
and consider the  complete set of two-loop EDM  graphs, shown in Fig.\
\ref{f1}, including those due  to CP-violating $a A_\mu Z_\lambda$ and
$H^+A_\mu W^-_\lambda$\footnote[1]{Analogous EDM contributions induced
by  charged-Higgs-boson  two-loop graphs  were  studied in  \cite{BCK}
within the context of three-Higgs-doublet models.}  couplings. At this
point, it is  worth stressing that the EDM  constraints we shall study
here  will have  important consequences  on Higgs-sector  CP violation
within the MSSM found recently \cite{APLB} and on related phenomena in
$B$-meson  decays,  dark-matter   searches  and  collider  experiments
\cite{CPHig}.

Our starting point  is the scalar  top and bottom mass matrices, which
may  conveniently   be expressed, in   the  weak basis ($\tilde{q}_L,\
\tilde{q}_R$), as follows:
\begin{equation}
  \label{Mscalar}
\widetilde{\cal M}^2_q \ =\ \left( \begin{array}{cc}
\tilde{M}^2_Q\, +\, m^2_q\, +\, \cos 2\beta M^2_Z\, ( T^q_z\, -\,
Q_q \sin^2\theta_w ) & m_q (A^*_q - \mu R_q )\\ 
m_q (A_q - \mu^* R_q) & \hspace{-0.2cm}
\tilde{M}^2_q\, +\, m^2_q\, +\, \cos 2\beta M^2_Z\, Q_q \sin^2\theta_w 
\end{array}\right)\, ,
\end{equation}
with $q=t,b$, $Q_t = 2/3$, $Q_b = -1/3$, $T^t_z = - T^b_z = 1/2$, $R_b
= \tan\beta = v_2/v_1$, $R_t  = \cot\beta$, and  $\sin\theta_w = (1  -
M^2_W/M^2_Z)^{1/2}$.  Moreover, $\tilde{M}^2_Q$    and $\tilde{M}^2_q$
are  soft-SUSY-breaking  masses for the  left-handed  and right-handed
scalar top and bottom quarks.  The matrix $\widetilde{\cal M}^2_q$ may
be  diagonalized through  a  unitary rotation,  which relates the weak
($\tilde{q}_L,\ \tilde{q}_R$)  to the mass eigenstates ($\tilde{q}_1,\ 
\tilde{q}_2$):
\begin{equation}
  \label{Rscalar}
\left( \begin{array}{c} \tilde{q}_L \\ \tilde{q}_R \end{array} \right)
\ =\ \left( \begin{array}{cc} 1 & 0 \\ 0 & e^{i\delta_q} \end{array} \right)
\left( \begin{array}{cc} \cos\theta_q & \sin\theta_q \\
          -\sin\theta_q & \cos\theta_q \end{array} \right)\,
\left( \begin{array}{c} \tilde{q}_1 \\ \tilde{q}_2 \end{array}\right)\, ,
\end{equation}
where $\delta_q  =  {\rm arg} (A_q  -  R_q \mu^*)$ and $\theta_q$  are
mixing angles determined by
\begin{eqnarray}
  \label{theta}
\cos\theta_q \!&=&\! \frac{m_q |A_q - R_q \mu^*|}{ 
\sqrt{m^2_q |A_q - R_q \mu^*|^2\, +\, [ (\widetilde{\cal M}^2_q)_{LL}
-  M^2_{\tilde{q}_1} ]^2} }\ ,\nonumber\\
\sin\theta_q \!&=&\! \frac{ |(\widetilde{\cal M}^2_q)_{LL}
-  M^2_{\tilde{q}_1}|}{ 
\sqrt{m^2_q |A_q - R_q \mu^*|^2\, +\, [ (\widetilde{\cal M}^2_q)_{LL}
-  M^2_{\tilde{q}_1} ]^2} }\ .
\end{eqnarray}
In Eq.\ (\ref{theta}), the quantity $(\widetilde{\cal M}^2_q)_{LL}$ is
the  $\{11\}$-matrix   element  of  the   scalar  quark   mass  matrix
$\widetilde{\cal  M}^2_q$.  In  addition,   the  mass  eigenvalues  of
$\widetilde{\cal M}^2_q$ are given by
\begin{eqnarray}
  \label{Mq12}
M^2_{\tilde{q}_1 (\tilde{q}_2)} & =& \frac{1}{2}\ \bigg\{
\tilde{M}^2_Q + \tilde{M}^2_q + 2m^2_q + T^q_z \cos 2\beta M^2_Z \nonumber\\
&& -(+)\ \sqrt{ \Big[ \tilde{M}^2_Q - \tilde{M}^2_q 
+ \cos 2\beta M^2_Z ( T^q_z - 2Q_q \sin^2\theta_w )\, \Big]^2\, 
                            +\, 4m^2_q |A^*_q -\mu R_q |^2 }\ \bigg\}.\quad
\end{eqnarray}

As usual, we  consider the convention in  which ${\rm arg}\, \mu$  and
${\arg}\,  A_{t,b}$ are the only physical  SUSY CP-violating angles in
the MSSM.  Then, CP      violation originates from    the  interaction
Lagrangian
\begin{equation}
  \label{Lcp}
{\cal L}_{\rm CP}\ =\ a \sum_{\tilde{q}=\tilde{t},\tilde{b}}\ \sum_{i,j=1,2}
\Big(\tilde{q}_i^*\, {\rm Re}\Gamma^{a\tilde{q}_i^*\tilde{q}_j}\,
\tilde{q}_j \Big)\ +\ 
\Big(\, H^+\! \sum_{i,j = 1,2}\,
\tilde{t}_i^*\, i {\rm Im} \Gamma^{H^+\tilde{t}_i^* \tilde{b}_j}\, \tilde{b}_j\
+\ {\rm H.c.}\Big)\,.
\end{equation}
The real and imaginary parts  of the couplings $a\tilde{q}^*\tilde{q}$
and   $H^+\tilde{t}_i^*  \tilde{b}_j$,      which  are   denoted    by
$\Gamma^{a\tilde{q}^*_i\tilde{q}_j}$ and     $\Gamma^{H^+\tilde{t}^*_i
  \tilde{b}_j}$, respectively, are given by
\begin{eqnarray}
  \label{ReHqq}
{\rm Re}\, \Gamma^{a\tilde{q}_i^*\tilde{q}_j} &=& v\,\xi_q\, 
\left( \begin{array}{cc} 1 & -\cot 2\theta_q \\ 
-\cot 2\theta_q & -1 \end{array} \right) ,\\
  \label{ImHtb}
{\rm Im}\, \Gamma^{H^+\tilde{t}_i^* \tilde{b}_j} &=& \frac{v}{\sqrt{2}}\,
\left( \begin{array}{cr}
-\,\frac{\cos\theta_t}{\cos\theta_b}\, \xi_b\, +\, 
\frac{\cos\theta_b}{\cos\theta_t}\, \xi_t & \quad 
\frac{\cos\theta_t}{\sin\theta_b}\, \xi_b\, +\, 
\frac{\sin\theta_b}{\cos\theta_t}\, \xi_t\\
-\,\frac{\sin\theta_t}{\cos\theta_b}\, \xi_b\, -\, 
\frac{\cos\theta_b}{\sin\theta_t}\, \xi_t &
\frac{\sin\theta_t}{\sin\theta_b}\, \xi_b\, -\, 
\frac{\sin\theta_b}{\sin\theta_t}\, \xi_t \end{array}\right)\nonumber\\
&&+\, \frac{\sqrt{2}\, m_b m_t \sin (\delta_b - \delta_t) }{
\sin\beta \cos\beta\, v^2}\, \left( \begin{array}{cc}
\sin\theta_t \sin\theta_b &  - \sin\theta_t \cos\theta_b \\
-\cos\theta_t \sin\theta_b & \cos\theta_t \cos\theta_b
\end{array}\right) \, ,
\end{eqnarray}
where $v =  \sqrt{v_1^2 + v^2_2}  = 2g_w/M_W$, and $\xi_t$ and $\xi_b$
are the CP-violating quantities
\begin{equation}
  \label{xiq} 
\xi_q\ =\  R_q\, \frac{\sin 2\theta_q m_q\, {\rm Im}\, ( \mu
             e^{i\delta_q})}{\sin\beta\, \cos\beta\, v^2}\ .\qquad
\end{equation}

%******************************************************************
%%%Figure 1
%******************************************************************

We shall first  calculate the couplings $a A_\mu  (k) A_\lambda (q) $,
$a g_\mu  (k) g_\lambda (q)$,  $a A_\mu  (k)  Z_\lambda (q)$  and $H^-
A_\mu (k)  W^+_\lambda (q)$,  which  are induced by  $\tilde{t}$-  and
$\tilde{b}$-mediated one-loop graphs shown in Fig.\ \ref{f1}. We adopt
the Feynman--'t Hooft gauge, and neglect Feynman diagrams that lead to
suppressed  EDM  contributions   proportional  to  $m^2_f$.   With the
convention  that the momenta   $q$ and $k$ flow   into the vertex, the
analytic  result  of  the  one-loop  couplings  is  found to  have the
gauge-invariant form
\begin{equation}
  \label{vertCP}
i\Gamma^{V}_{\mu\lambda}(k,q)\ =\ i A^{V}(q^2)\, \Big[\, (q\cdot k)\,
g_{\mu\lambda}\: -\:  q_\mu k_\lambda \, \Big]\, ,
\end{equation}
where the  superscript $V$ denotes the  gauge boson propagating in the
loop, and
\begin{eqnarray}
  \label{ABgamma}
A^\gamma (q^2) \!&=&\! \sum_{q=t,b}
\frac{N_c e^2 Q^2_q v}{8\pi^2}\sum_{i=1,2} 
(-1)^{i+1}\xi_q \int_0^1 dx\, \frac{x(1-x) }{M^2_{\tilde{q}_i}\,
-\, q^2 x(1-x)}\ ,\\
  \label{ABZ}
A^Z (q^2) \!&=&\! \sum_{q=t,b}
\frac{N_c g_w e Q_q}{16\cos\theta_w\, \pi^2}\sum_{i,j=1,2}  
      {\rm Re}\, \Gamma^{a\tilde{q}_i^*\tilde{q}_j}
K^q_{ij}\nonumber\\
&&\times\, \int_0^1 dx\, \frac{x(1-x) }{M^2_{\tilde{q}_i}x\, +\, 
M^2_{\tilde{q}_j} (1-x)\, -\, q^2 x(1-x)}\ ,\\
  \label{AW} 
A^W (q^2) \!&=&\! \frac{N_c g_w e}{8\sqrt{2} \pi^2}
\sum_{i,j = 1,2} \Big(\Gamma^{H^+\tilde{t}_i^*\tilde{b}_j}\Big)^*\,
K^{tb}_{ij}\, \int_0^1 dx\, \frac{x(1-x)\, [ Q_t x\, +\,
Q_b (1-x)]}{M^2_{\tilde{t}_i} x\, +\, M^2_{\tilde{b}_j} (1-x)\,
-\, q^2 x(1-x)}\ .
\end{eqnarray}
The  corresponding  one-loop form  factor  $A^g$  may  be obtained  by
replacing the  colour factor $N_c  =3$ by $1/2$,  and $Q^2_q$ by  1 in
Eq.\  (\ref{ABgamma}).   In Eqs.\  (\ref{ABZ})  and (\ref{AW}),  $K^q$
($q=t,b$)   and  $K^{tb}$   are  $(2\times   2)$-non-unitary  matrices
describing  the   mixing  in  the   $Z\tilde{q}^*_i  \tilde{q}_j$  and
$W^+\tilde{t}^*_i \tilde{b}_j$ sectors, respectively:
\begin{eqnarray}
  \label{Kq}  
K^q &=& \left( \begin{array}{cc}
2T^q_z \cos^2\theta_q - 2 Q_q\sin^2\theta_w & T^q_z \sin 2\theta_q \\
T^q_z \sin 2\theta_q & 2T^q_z \sin^2\theta_q - 2 Q_q\sin^2\theta_w
\end{array} \right),\\
  \label{Ktb}
K^{tb} & =& \left( \begin{array}{cc}
\cos\theta_t \cos\theta_b & \cos\theta_t \sin\theta_b \\
\sin\theta_t \cos\theta_b & \sin\theta_t \sin\theta_b
\end{array}\right)\, .
\end{eqnarray}
At this  stage,  we should  remark that  there are  also contributions
originating  from chargino loops.    However, these  contributions are
proportional to ${\rm arg}\mu$, on which strict constraints exist from
one-loop graphs that contribute to the electron EDM, and may therefore
be   neglected.     Finally,  we should notice      that  the one-loop
$H^-AW^+$-coupling   receives   its   gauge-invariant  form  of   Eq.\
(\ref{vertCP}) in   the Feynman--'t Hooft  gauge,  after including the
$H^-W^+$ and $H^-G^+$  wave   functions  and the   respective  tadpole
contribution related to the  $H^-G^+$ transition,  as shown in  Figs.\
\ref{f1}(e)--(g).

It  is now straightforward to compute  the individual contributions to
the EDM  of a  fermion that come  from  quantum corrections  involving
$\gamma$,  $Z$ and  $W^\pm$ bosons  in the loop.  These individual EDM
contributions shown in  Fig.\ \ref{f1} may  conveniently  be cast into
the form:
\begin{eqnarray} 
  \label{EDMfgamma}   
\bigg( \frac{d_f}{e}\bigg)^\gamma &=& Q_f\,   
\frac{N_c \alpha}{32\pi^3}\, \frac{\tan\beta\ m_f}{M^2_a}\ \sum_{q = t,b}\ 
\xi_q\, Q^2_q\,\bigg[\, F\bigg(\frac{M^2_{\tilde{q}_1}}{M^2_a}\bigg)\ -\ 
F\bigg(\frac{M^2_{\tilde{q}_2}}{M^2_a}\bigg)\, \bigg]\, ,\\
   \label{EDMfZ}
\bigg( \frac{d_f}{e}\bigg)^Z &=& -(T^f_z - 2Q_f \sin^2\theta_w)\,   
\frac{N_c \alpha_w}{64\cos^2\theta_w\, \pi^3}\, 
\frac{\tan\beta\ m_f}{M^2_a}\nonumber\\
&&\times \sum_{q = t,b}\ 
\sum_{i,j=1,2}\ \frac{1}{v}\ {\rm Re}\,  
\Gamma^{a\tilde{q}_i^*\tilde{q}_j}\, K^q_{ij}\, Q_q\
G\bigg( \frac{M^2_Z}{M^2_a},\ \frac{M^2_{\tilde{q}_i}}{M^2_a},\ 
\frac{M^2_{\tilde{q}_j}}{M^2_a}\bigg)\, ,\\ 
   \label{EDMfW}
\bigg( \frac{d_f}{e}\bigg)^W &=&    
\frac{N_c \alpha_w}{128\sqrt{2} \pi^3}\, 
\frac{\tan\beta\ m_f}{M^2_{H^+}}\,  
\sum_{i,j=1,2}\ \frac{1}{v}\ {\rm Im}\,  
\Gamma^{H^+\tilde{t}_i^*\tilde{b}_j}\, K^{tb}_{ij}\nonumber\\
&&\times\,
\bigg[\,Q_t\, G\bigg(\frac{M^2_W}{M^2_{H^+}},\ 
\frac{M^2_{\tilde{t}_i}}{M^2_{H^+}},\ 
\frac{M^2_{\tilde{b}_j}}{M^2_{H^+}}\bigg)\,
+\, Q_b\, G\bigg(\frac{M^2_W}{M^2_{H^+}},\ 
\frac{M^2_{\tilde{b}_i}}{M^2_{H^+}},\ 
\frac{M^2_{\tilde{t}_j}}{M^2_{H^+}}\bigg)\, \bigg]\,,
\end{eqnarray}
where  $\alpha  = e^2/(4\pi)$  and $\alpha_w =   g^2_w/(4\pi)$ are the
electromagnetic and weak  fine structure  constants, respectively, and
$F(z)$ and $G(a,b,c)$ are two-loop functions given by
\begin{eqnarray} 
  \label{Fz}
F(z) &=& \int_0^{1} dx\ \frac{x(1-x)}{z - x(1-x)}\ 
\ln \bigg[\,\frac{x(1-x)}{z}\,\bigg]\, ,\\
G(a,b,c) &=& \int_0^1 dx\ x\, \bigg[\, \frac{a x(1-x) \ln a}{
(a-1) [a x(1-x) - bx - c (1-x)]}\nonumber\\ 
&&\hspace{-2cm}+\ \frac{ x (1-x) [ b x + c (1-x)]}{
[a x(1-x) - bx - c(1-x)]\, [x(1-x) - bx - c(1-x)] }\
\ln \bigg( \frac{ b x + c (1-x)}{ x (1-x) } \bigg)\, \bigg]\, .\quad
\end{eqnarray}
Note that  $2G(0,a,a) = - F(a)$.   Among the EDM terms  given by Eqs.\
(\ref{EDMfgamma})--(\ref{EDMfW}),  the  fermion  EDM  induced  by  the
photon-exchange   graphs  $(d_f/e)^\gamma$  represents   the  dominant
contribution  \cite{CKP}.   For  completeness,  we give  the  two-loop
contribution to the CEDM of a coloured fermion \cite{CKP}
\begin{equation} 
  \label{CEDMf}   
\bigg( \frac{d^C_f}{e}\bigg)^g \ =\   
\frac{\alpha_s}{64\pi^3}\, \frac{\tan\beta\ m_f}{M^2_a}\ \sum_{q = t,b}\ 
\xi_q\, \bigg[\, F\bigg( \frac{M^2_{\tilde{q}_1}}{M^2_a}\bigg)\ -\ 
F\bigg(\frac{M^2_{\tilde{q}_2}}{M^2_a}\bigg)\, \bigg]\, .
\end{equation}

We can  now estimate the neutron EDM  $d_n$ induced by $d_u$ and $d_d$
in the valence quark model,  by including QCD renormalization  effects
\cite{CKP}, i.e.\
\begin{equation}
  \label{dne}
\frac{d_n}{e}\ \approx\ 
\bigg(\,\frac{g_s (M_Z )}{g_s (\Lambda )}\,\bigg)^{32/23}\
\bigg[\, \frac{4}{3}\,\bigg(\frac{d_d}{e}\bigg)_\Lambda\ -\ 
\frac{1}{3}\,\bigg(\,\frac{d_u}{e}\,\bigg)_\Lambda
\bigg]\, .
\end{equation}
The $u$- and $d$-quark masses occurring  in Eq.\ (\ref{EDMfgamma}) are
running masses evaluated  at the low-energy  hadronic scale $\Lambda$. 
In Eq.\  (\ref{dne}),  we have assumed  that the renormalization-group
running  factors of the strong coupling  constant $g_s$  from $m_b$ to
$\Lambda$ is  almost  of order 1.    To be specific,  we  consider the
values: $m_u (\Lambda ) = 7$ MeV, $m_d (\Lambda ) = 10$ MeV, $\alpha_s
(M_Z) = 0.12$, and $g_s (\Lambda  )/(4\pi) = 1/\sqrt{6}$ \cite{DDLPD}. 
Likewise, the light-quark CEDMs $d_u^C$ and $d^C_d$  lead to a neutron
EDM
\begin{equation}
  \label{cdne}
\frac{d^C_n}{e}\ \approx\ 
\bigg(\,\frac{g_s (M_Z )}{g_s (\Lambda )}\,\bigg)^{74/23}\
\bigg[\,\frac{4}{9}\,\bigg(\frac{d^C_d}{e}\bigg)_\Lambda\ +\
\frac{2}{9}\,\bigg(\frac{d^C_u}{e}\bigg)_\Lambda\, \bigg],
\end{equation}
where  in turn  the strong coupling  constant  $g_s$ and the  $u$- and
$d$-quark  masses in $d^C_u$ and $d^C_d$  are calculated  at the scale
$\Lambda$.

\begin{figure}
   \leavevmode
 \begin{center}
   \epsfxsize=14.0cm
    \epsffile[0 0 539 652]{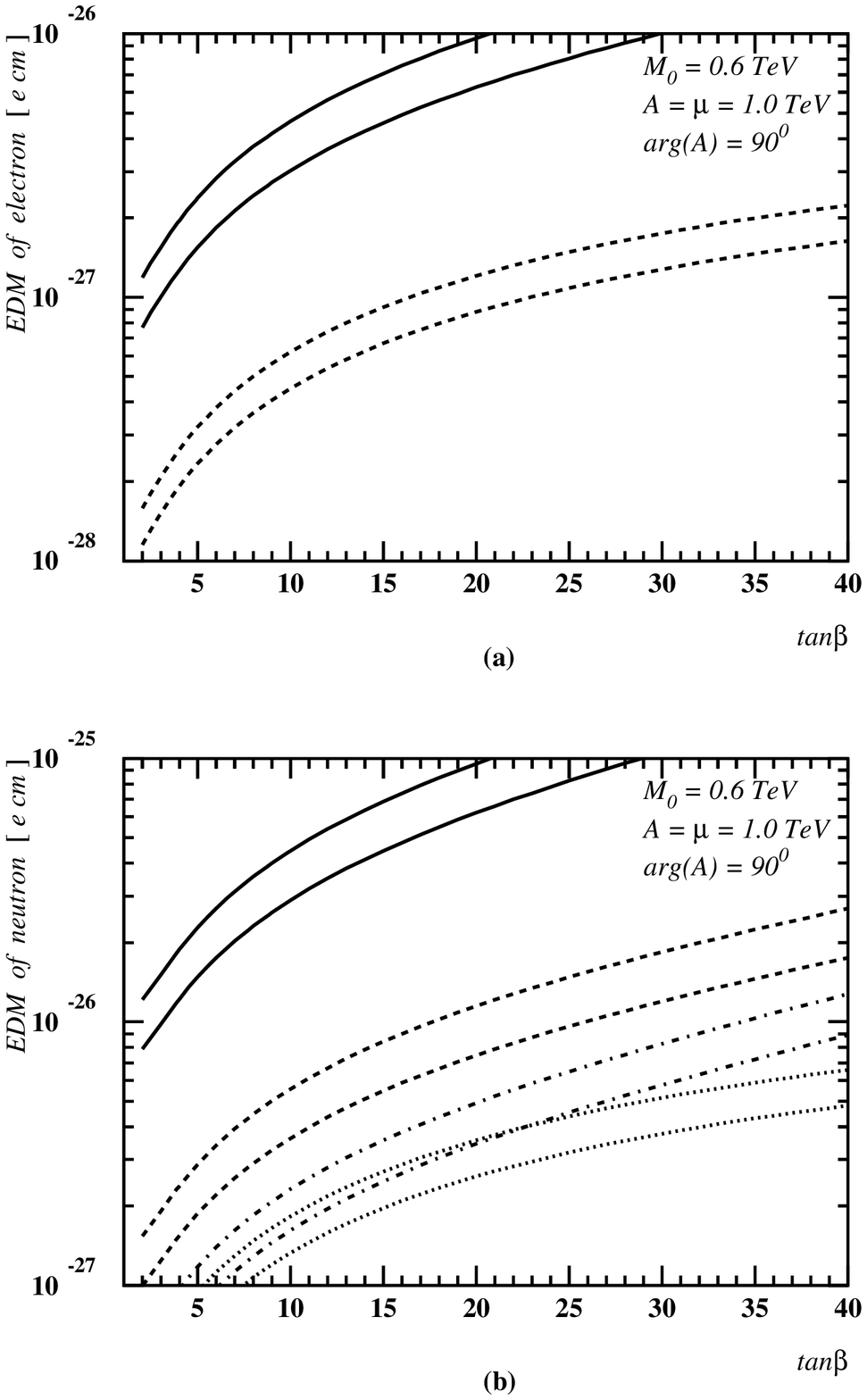}
 \end{center}
 \vspace{-0.5cm} 
\caption{Numerical estimates of the individual two-loop EDM
contributions as a function of $\tan\beta$: {\bf (a)} $(d_e)^\gamma$
(solid line), $(d_e)^W$ (dashed line); {\bf (b)} $-(d_n)^C$ (solid
line), $(d_n)^\gamma$ (dashed line), $(d_n)^W$ (dotted line),
$(d_n)^Z$ (dash-dotted line).  Lines of the same type from the upper
to the lower one correspond to $M_a = 150$ and 300 GeV,
respectively.}\label{f2}
\end{figure}

\begin{figure}
   \leavevmode
 \begin{center}
   \epsfxsize=14.0cm
    \epsffile[0 0 539 652]{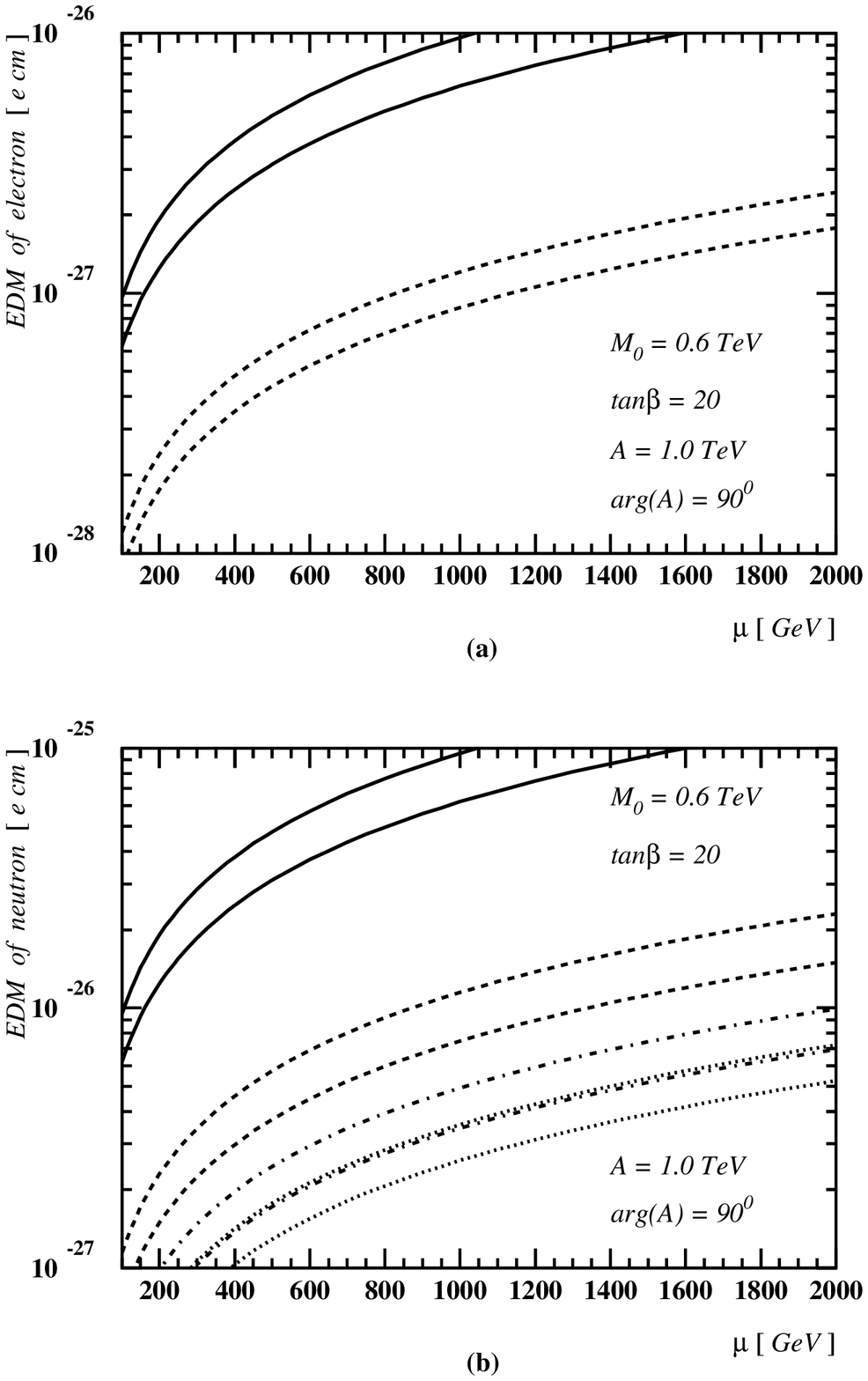}
 \end{center}
 \vspace{-0.5cm} 
\caption{Numerical estimates of the individual two-loop EDM
contributions as a function of $\mu$: {\bf (a)} $(d_e)^\gamma$
(solid line), $(d_e)^W$ (dashed line); {\bf (b)} $-(d_n)^C$ (solid
line), $(d_n)^\gamma$ (dashed line), $(d_n)^W$ (dotted line),
$(d_n)^Z$ (dash-dotted line).  Lines of the same type from the upper
to the lower one correspond to $M_a = 150$ and 300 GeV,
respectively.}\label{f3}
\end{figure}

In the following, we shall  give a more quantitative discussion of the
individual  two-loop contributions  to electron  and neutron  EDMs for
$M_a = 150$  and 300 GeV.  At the  tree level, the charged-Higgs-boson
mass $M_{H^+}$ is related to the would-be CP-odd mass $M_a$ by
\begin{equation}
  \label{MaMHP}
M^2_{H^+}\ =\ M^2_a\, +\, M^2_W\, .
\end{equation}
Even   though this very  last  relation receives appreciable radiative
corrections  in the MSSM  with Higgs-sector CP violation \cite{MHP}, we
shall still   make  use  of  Eq.\   (\ref{MaMHP}),  as required  by  a
consistent expansion  in perturbation theory.   As we  have explicitly
demonstrated in \cite{CKP},  the  EDMs crucially depend on  $\mu$  and
$\tan\beta$ through  $\xi_{t,b}$  in Eq.\ (\ref{xiq})  and through the
couplings of $a$ and $H^+$ to electron and $d$ quark.  For the purpose
of illustration, we therefore plot in Figs.\ \ref{f2} and \ref{f3} the
numerical predictions for  electron and neutron  EDMs  as functions of
$\tan\beta$ and   $\mu$, respectively.  Specifically,  we consider the
following values for the SUSY parameters:
\begin{eqnarray}
  \label{par}
{\rm Fig.\ \ref{f2}:} && M_0\, =\, \tilde{M}_Q\, =\, \tilde{M}_u\, =\, 
\tilde{M}_d\, =\, 0.6\
{\rm TeV}, \quad
A\, =\, |A_t|\, =\, |A_b|\, =\, 1\ {\rm TeV},\nonumber\\
&&A\, =\, \mu\, =\, 1\ 
{\rm TeV}\,,\quad
{\rm arg}\,A\, =\, 90^\circ\nonumber\\ 
{\rm Fig.\ \ref{f3}:} && M_0\, =\, \tilde{M}_Q\, =\, \tilde{M}_u\, =\, 
\tilde{M}_d\, =\, 0.6\ {\rm TeV}, \quad
A\, =\, |A_t|\, =\, |A_b|\, =\, 1\ {\rm TeV},\nonumber\\
&&\tan\beta\, =\, 20\,,\quad
{\rm arg}\,A\, =\, 90^\circ
\end{eqnarray}
In addition, we assume that  the $\mu$-parameter is real. In agreement
with \cite{CKP}, we find that  the dominant EDM effects originate from
the $a  A_\mu A_\lambda$-coupling  in $d_e$, $(d_e)^\gamma$,  and from
the  $a  g_\mu g_\lambda$-coupling  in  $d_n$,  $(d_n)^C$.  In  Figs.\ 
\ref{f2}(a)  and  \ref{f3}(a),  we  see that  the  charged-Higgs-boson
two-loop   contribution   to   EDM,   $(d_e)^W$,   is   smaller   than
$(d_e)^\gamma$ by a  factor 8.  On the other  hand, Figs.\ \ref{f2}(b)
and  \ref{f3}(b)  show   that  the  corresponding  EDM  contributions,
$(d_n)^W$, $(d_n)^\gamma$, and $(d_n)^Z$,  are all of comparable size. 
They are roughly one order of magnitude smaller than $(d_n)^C$, and of
opposite sign.

In conclusion, we have shown  that EDM constraints on the CP-violating
parameters related to the sectors of  scalar top and bottom quarks can
be significant  for  $\tan\beta \stackrel{>}{{}_\sim} 10$,  and $\mu,\ 
A_{t,b} \stackrel{>}{{}_\sim} 0.5$ TeV. In particular, we have studied
additional  two-loop contributions  mediated  by   $W$- and  $Z$-boson
interactions, which are found to  be sub-dominant but  non-negligible,
and are   therefore  expected to  play  an important  role  in  future
phenomenological analyses of Higgs-sector CP violation in the MSSM.

\subsection*{Note added}
While revising  the paper,  I became aware  of Ref.\  \cite{CCK} which
addresses  the same  topic.  After  the final  revisions,  the results
obtained by  the two groups  agree both analytically  and numerically.
The  author  wishes  to  thank  Darwin Chang  and  Wai-Yee  Keung  for
discussions.

%******************************************************************
%%%Figure 2
%******************************************************************

%******************************************************************
%%%Figure 3
%******************************************************************

\end{document}